\documentclass[preprint2]{aastex}
\usepackage[]{natbib}
\usepackage[]{graphicx}

\shorttitle{HM~Sge}
\shortauthors{S. P. S. Eyres et al.}

\volume{551}

\begin{document}
\title{The inner nebula and central binary of the symbiotic star
HM~Sge} \author{ S. P. S. Eyres and
M. F. Bode} \affil{Astrophysics Research Institute, Liverpool John
Moores University, Twelve Quays House, Egerton Wharf, Birkenhead, CH41
1LD, UK} \email{spse@astro.livjm.ac.uk, mfb@astro.livjm.ac.uk}

\author{A. R. Taylor}
\affil{The Department of Physics and Astronomy,
The University of Calgary,
2500 University Dr. N.W.,
Calgary,
Alberta,
T2N 1N4,
Canada}
\email{russ@ras.ucalgary.ca}

\and

\author{M. M. Crocker, R. J. Davis}
\affil{University of Manchester, Jodrell Bank Observatory, Macclesfield, Cheshire, SK11 9DL, UK}
\email{mc@jb.man.ac.uk, rjd@jb.man.ac.uk}

\begin{abstract}
We present contemporaneous {\em HST} WFPC2 and VLA observations of the
symbiotic nova HM~Sge.  We identify a number of discreet features at
spatial scales smaller than $\sim$~0.1~arcsec embedded in the extended
nebula, with radio and optical emission well correlated in the inner
1~arcsec. For the first time we measure the positions of the binary
components of a symbiotic star directly. We estimate the projected
angular binary separation to be 40$\pm$9~milli--arcsec, with the
binary axis at position angle 130$^{\circ}\pm$10$^{\circ}$.  The
latter is consistent with previous estimates made by indirect methods.
The binary separation is consistent with a previous estimate of 50~au
if the distance is 1250$\pm$280~pc. Temperature and density
diagnostics show two distinct regions in the surrounding nebula, with
a cool wedge to the south--west. An extinction map indicates the true
interstellar extinction to be no more than E(B--V) = 0.35.  This is
consistent with a minimum distance of $\sim$700~pc, but this would be
reduced if there is a circumstellar contribution to the minimum in the
extinction map. The extinction map also suggests a patchy dust
distribution.  We suggest that a southern concentration of dust and
the south--west wedge are associated with the cool component wind.
Alternatively, the southern dust concentration is the cause of the
cool wedge, as it shields part of the nebula from the hot component
radiation field.

%\vspace*{6cm}
\end{abstract}
\keywords{binaries: symbiotic -- stars: individual (HM~Sge) --
circumstellar matter -- radio continuum: stars}

\section{Introduction}
\label{sec-intro}

Symbiotic stars are an extreme case of interacting binaries, with
separations of a few to a few 10s of au.  They comprise a cool
component (CC), typically a red giant or Mira--type variable, and a
hot component (HC), usually a white dwarf associated with an ionized
component of the CC wind.  An infrared sub--classification has been made,
dividing the class into D(usty)--types, and S(tellar)--types, based on
the dominant contribution to the IR.  Early modeling of the radio
emission has shown that D--types typically have separations of 10 or
more times that of S--types.  Some of these objects show very slow
optical outbursts, similar to those of novae, but lasting decades (see
e.g. \citealt{Kenyon86} and \citealt{Mikolajewska97} for further
discussion.)

HM~Sge is a D--type symbiotic star which underwent such an optical
outburst in 1975. Since then, there has been evidence of dust
obscuration events, as in 1985 \citep{Munari89a} when A$_V \simeq$
13.5 \citep{Whitelock88a,Munari89a}, and in 1979--1980 when A$_V
\simeq$ 12 \citep{Thronson81}. The different values of A$_V$ have led
to a suggestion of clumping in the dust \citep[e.g.][]{Thronson81},
which was also suggested by \citet{Richards99} to explain the nebula
structural changes seen in the radio. Various distance estimates to
HM~Sge are summarized by \citet{Ivison91}.  Those authors' own
estimate of 1300~pc was based on a reddening estimate of E(B--V) =
0.53.  \citet{Kenyon88} used far--infrared IRAS colors to derive the
extinction and a distance of 1800~pc.  These estimates clearly depend
on the effects of variable circumstellar extinction.  Recently the
Mira--type cool component has been clearly seen in
spectro--polarimetry \citep{Schmid00a}, indicating that dust
obscuration has subsided somewhat.  The spectra of \citet{Schmid00a}
are dominated by the line emission, with a rising continuum beyond
$\sim$8000~\AA.  They also suggested that the binary axis had a
position angle of 123$^\circ$ in 1998.  This contradicts tentative
conclusions by \citet{Richards99} placing the Mira--type due north of
the hot component (i.e. position angle $\sim 0^\circ$).  Many of the
observations of the outburst and subsequent development of HM~Sge are
described in \citet{Nussbaumer90} and \citet{Murset94}.

\citet{Corradi99} have conducted ground--based imaging of the extended
nebula of HM~Sge.  They find a number of filaments and discrete
features embedded in a nebula extending to $\sim 13$~arcsec.  This
emission is extremely faint, and if the nebula is due to activity in
the central binary, suggests a much longer history to such activity
than observed since 1975.

HM~Sge was detected in the radio shortly after the optical outburst
\citep{Feldman77}. Radio surveys showed HM~Sge to be one of the
brightest radio emitters in the class \citep{Seaquist84a,Seaquist93a}.
\citet{Kwok84} found a diffuse halo $\sim$0.5~arcsec in diameter,
surrounding a central nebula $\sim$0.15~arcsec in size.  More
recently, \citet{Eyres95} demonstrated that the structure seen in the
radio could be correlated with {\em HST} images at the same spatial
resolution, suggesting that at least the HC was centrally located in
the inner nebula.  \citet{Richards99} traced the development of the
radio structure over 5 years, and related it to the binary motion.
They derived a period of 80~years and a binary separation of
50~au\footnote{The references in \citet{Richards99} to a binary
separation of 25~au are made in error -- the binary orbital
semi--major axis is 25~au, giving the binary separation as 50~au.}.
Non--thermal emission associated with an E--W outflow has also been
found \citep{Eyres95}, and \citet{Richards99} showed that this is rare
in symbiotic stars due to the relatively short timescale on which the
emission dissipates.

Here, we discuss our observations of the nebula associated with this
star using WFPC2 on the {\em Hubble Space Telescope} {\em
(HST)}. We also present contemporaneous radio images of the inner
nebula made with the Very Large Array (VLA). These observations allow
us to determine the physical parameters in the nebula, and to attempt
to relate the conditions with the binary interaction and the outburst
history of the system.

\section{Observations}
\label{sec-obs}

The {\em HST} observations were made on 1999~October~22, as part of a
GO program 8330 on symbiotic stars, as shown in Table~\ref{tab-log}.
Three orbits were allocated to HM~Sge, and observations were made in
seven filters, including F218W, F437N, F469N, F487N, F502N, F547M and
F656N.
%Both ``long'' and
%``short'' exposures were made, to allow for the relative brightness of
%the central star and the extended nebula.  
The exposure times and dominant lines for images presented here
%each target in each filter 
are given in Table~\ref{tab-log}. Further details of these filters are available from \citet{Biretta96}.
 %Where
%observing time allowed, each exposure was split into two
%sub--exposures to allow for cosmic ray subtraction, and for most short
%exposures, the two sub--exposures were also dithered to allow the
%recovery of the full spatial resolution.  Which exposures were
%dithered is noted in the individual figure captions, and in
%section~\ref{sec-results}.
\begin{table*}
\begin{center}
\caption[]{Observation log.\label{tab-log}}
\begin{tabular}{lccccc}
\tableline\tableline
Filter & Exposure &
$\bar{\lambda}$ & $\Delta\bar{\lambda}$ & Peak $\lambda$ & Scientific features\\
&  times (s)& (\AA) & (\AA) & (\AA) & and wavelengths (\AA)\\
\tableline
F218W\tablenotemark{\dag}
	& 40  & 2136 & 355.9 & 2091 & Interstellar absorption\\
F218W	& 100  & 2136 & 355.9 & 2091 & Interstellar absorption\\
F437N	& 800 & 4369 & 25.2  & 4368 & [O III] $\lambda$4363\\
F469N	& 100 & 4695 & 24.9  & 4699 & He~II $\lambda$4686\\
F487N	& 200 & 4865 & 25.8  & 4863 & H$\beta~\lambda$4861 \\
F502N	& 100 & 5012 & 26.8  & 5009 & [O III] $\lambda\lambda$4959, 5007 \\
F547M\tablenotemark{\dag}
	& 2   & 5454 & 486.6 & 5362 & Str\"{o}mgren y\\
F547M	& 20  & 5454 & 486.6 & 5362 & Str\"{o}mgren y\\
F656N\tablenotemark{\dag}
	& 100 & 6562 & 22.0 & 6561 & H$\alpha~\lambda$6563 \\
\tableline
\end{tabular}
\tablenotetext{\dag}{dithered images}
\end{center}
\end{table*}
The calibrated data were retrieved from the {\em HST} archive.  Each
exposure was executed in two sub--exposures to allow cosmic ray
subtraction.  The F656N image presented here was also dithered to
allow recovery of the full spatial resolution \citep{Biretta96}.  In
addition, the stellar positions (section~\ref{ssec-positions}) were
determined from dithered images taken through the F218W and F547M
filters.  The pixel size was 0.0455~arcsec for the undithered images
and 0.02275~arcsec for the dithered ones.

The VLA observations were made on 1999~September~26 at 8.56~GHz and
23~GHz. Comparison with the primary calibrator 1331+305 (3C286) gave
the flux of secondary calibrator 1935+205 as 0.385$\pm$0.001~Jy at
8.56~GHz and 0.41$\pm$0.01~Jy at 23~GHz.  The complex gain solutions
for this calibrator were applied to HM~Sge. MERLIN observations by
\citet{Richards99} show that the structures in the inner nebula move
by $\sim$4~milli--arcsec per year and the flux density remained
roughly constant at 22~GHz between 1994 and 1996. Thus the VLA and
{\em HST} images can be directly compared.

\section{Results}
\label{sec-results}

\begin{figure*}
%\epsscale{0.9617} % to ensure same scale as fig\ref{fig-radio}
%    \plotone{HST1.eps} 
\includegraphics[width=170mm, scale=0.9617]{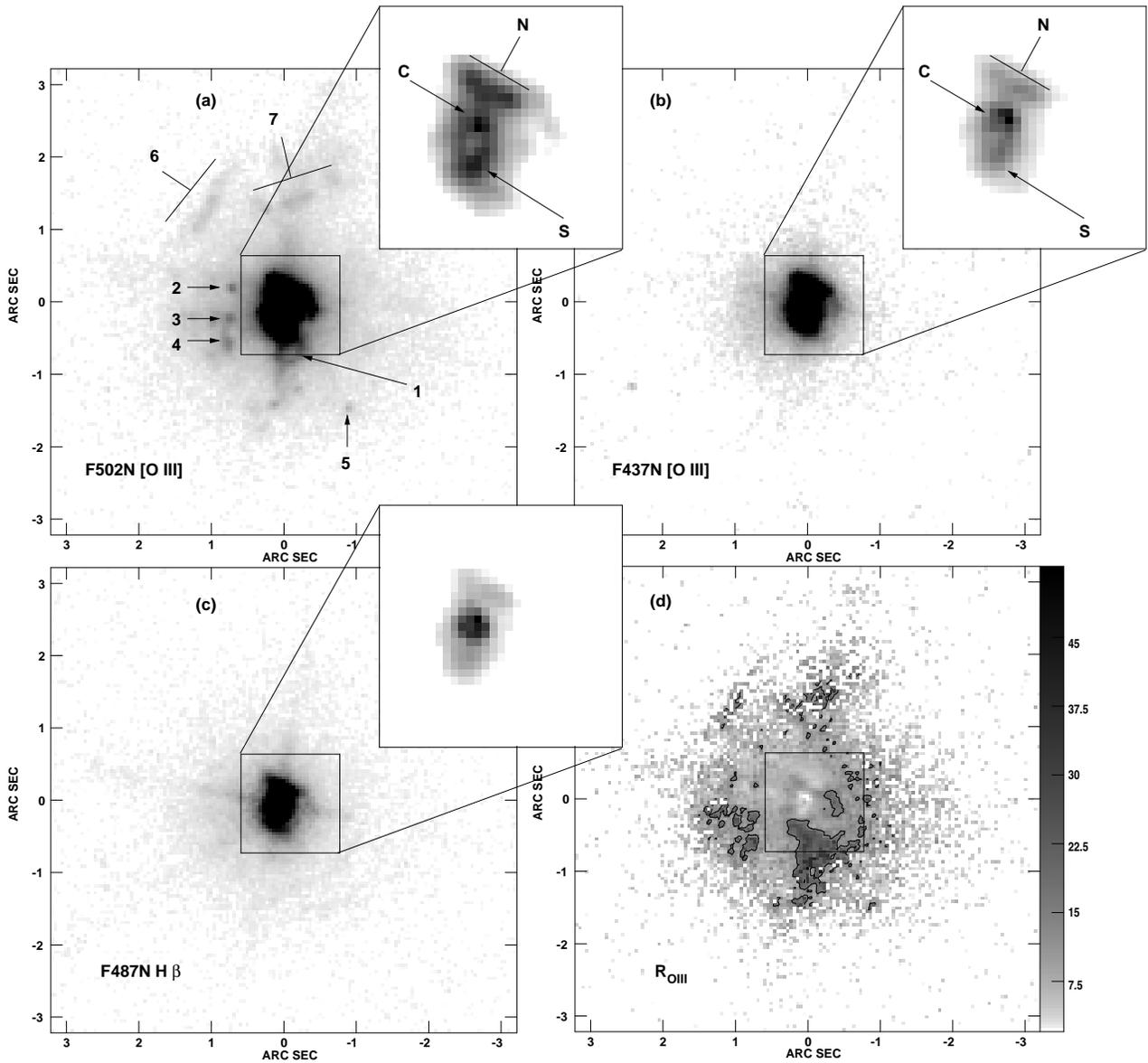}
\caption{{\em HST} images of HM Sge, dereddenned for
E(B--V) = 0.35, in the WFPC2 filters (a) F502N; greyscale range
9.1$\times$10$^{-15}$ to 1.3$\times$10$^{-12}$ (main panel) and
1.3$\times$10$^{-12}$ to 2.9$\times$10$^{-11}$ (inset); (b) F437N;
greyscale range 4.7$\times$10$^{-15}$ to 2.3$\times$10$^{-13}$ (main
panel) and 2.3$\times$10$^{-13}$ to 10$^{-11}$ (inset); and (c) F487N;
greyscale range 3.8$\times$10$^{-15}$ to 9.6$\times$10$^{-13}$ (main
panel) and 9.6$\times$10$^{-13}$ to 2.5$\times$10$^{-11}$ (inset).
Units are erg~s$^{-1}$~cm$^{-2}$~\AA$^{-1}$~arcsec$^{-2}$. In each
case the inset is at the same scale as used in the radio images in
Figs.~\ref{fig-radio}(a) and (b).  Image (d) is the ratio of images
(a) and (b) with greyscale range 2 to 50, and a single contour at
$R_{OIII}$ = 15. Uncertainties in $R_{OIII}$ are of order 10\%.
\label{fig-HST1}}
\end{figure*}
\begin{figure*}
%\epsscale{0.8858} % to ensure same scale as fig\ref{fig-radio}
%    \plotone{HST2.eps} 
\includegraphics[width=170mm, scale=0.8858]{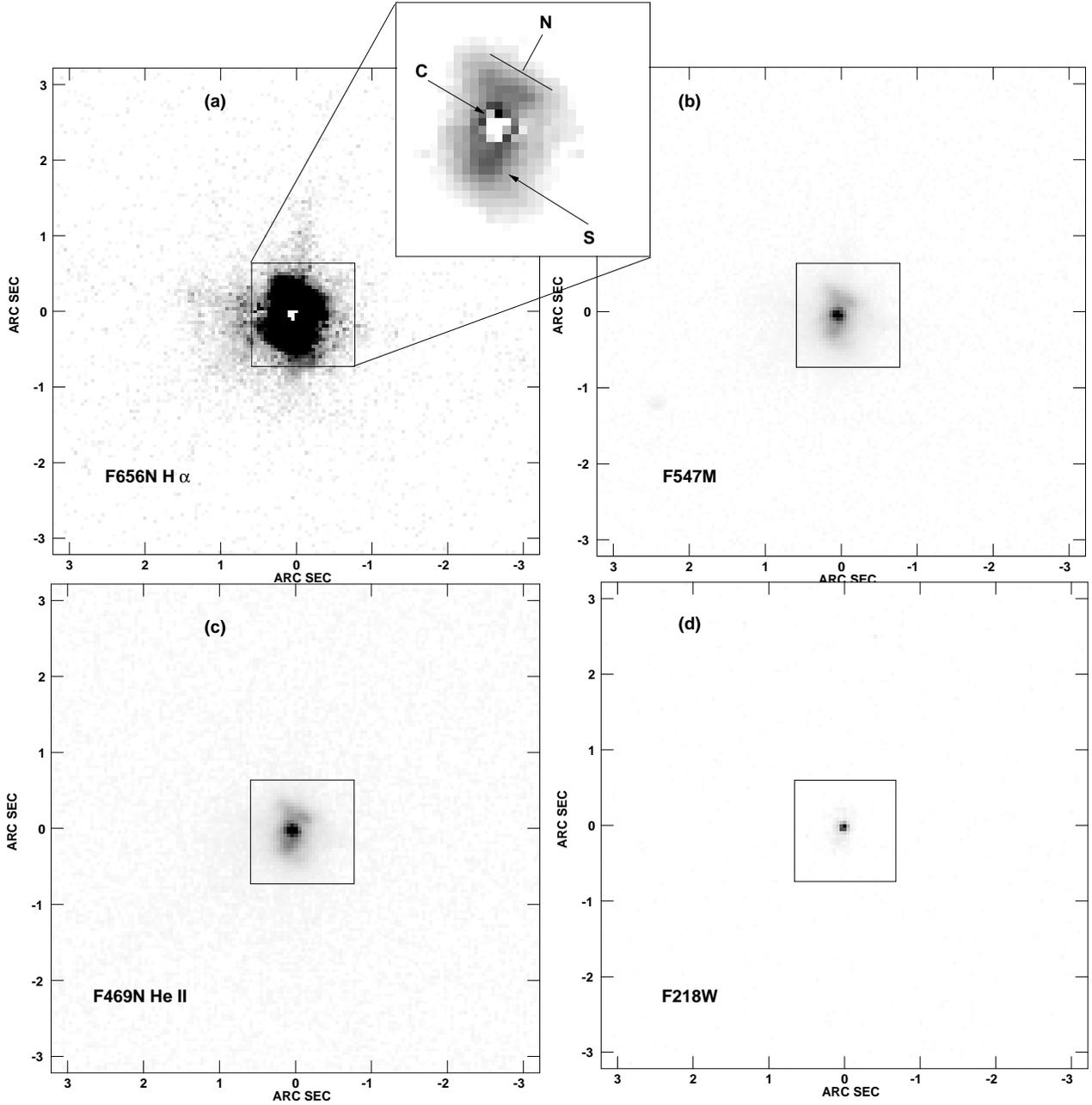}
\caption{{\em HST} images of HM Sge,
dereddenned for E(B--V) = 0.35, in the WFPC2 filters (a) F656N star
subtracted; greyscale range 1.47$\times$10$^{-15}$ to
2.93$\times$10$^{-14}$ (main panel) and 2.93$\times$10$^{-14}$ to
2.14$\times$10$^{-12}$ [inset, same scale as Figs.~\ref{fig-radio}(a)
and (b)]; (b) F547M; greyscale range 1.58$\times$10$^{-16}$ to
4.74$\times$10$^{-13}$; (c) F469N; greyscale range
2.09$\times$10$^{-16}$ to 1.47$\times$10$^{-11}$; and (d) F218W;
greyscale range 2$\times$10$^{-13}$ to 2$\times$10$^{-12}$. Units are
erg~s$^{-1}$~cm$^{-2}$~\AA$^{-1}$~arcsec$^{-2}$. In each case the
inset is at the same scale as used in the radio images in
Fig.~\ref{fig-radio}.
\label{fig-HST2}}
\end{figure*}

\begin{figure*}
%    \plotone{radio.eps} 
\includegraphics[width=170mm]{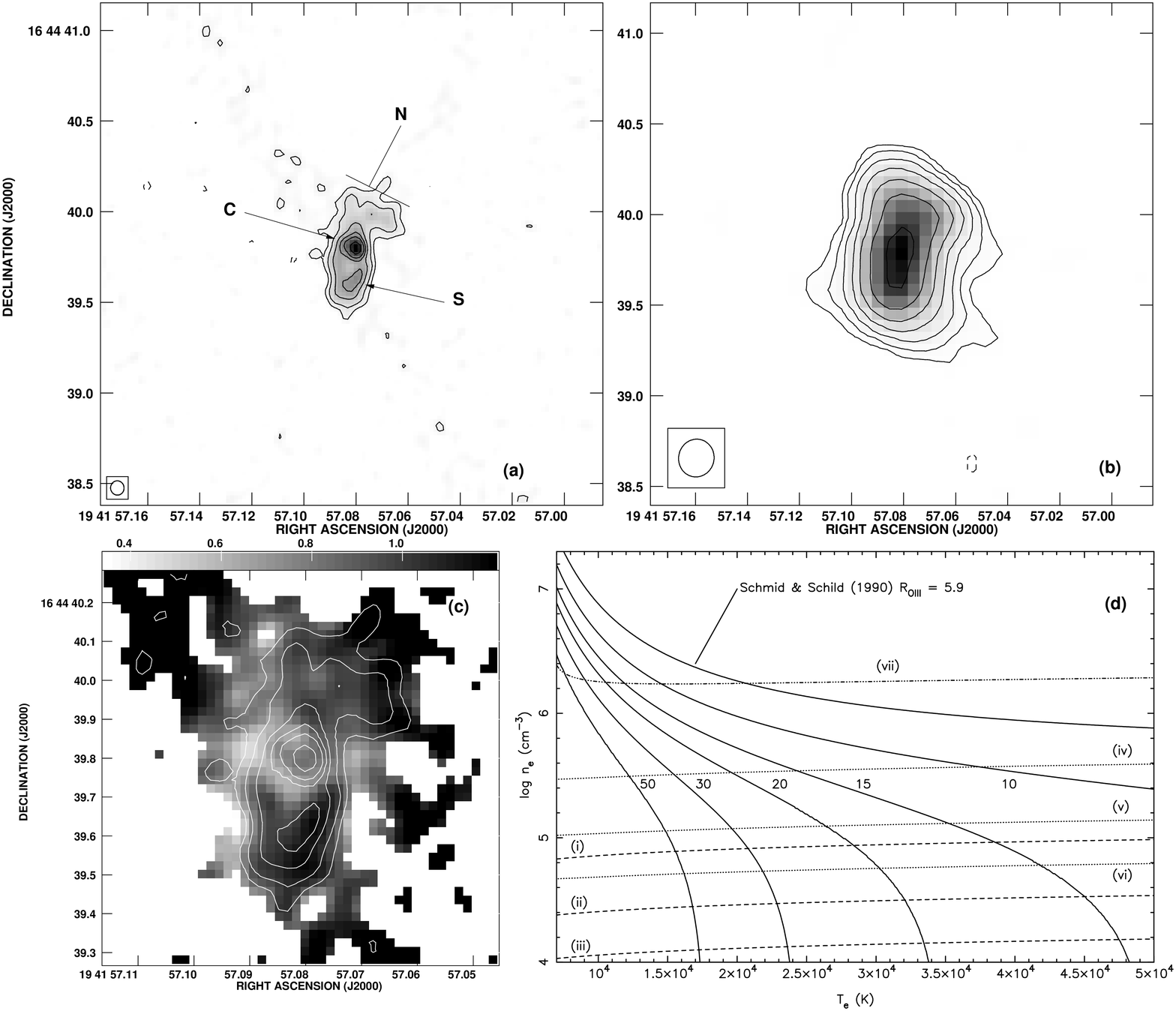}
	\caption{Radio images of HM Sge at (a) 23~GHz with the
VLA; contours are $-$3, 3, 6, 12, 18, 24 and 36 $\times$
262~$\mu$Jy~beam$^{-1}$, greyscale range is 0.262 to
12.05~mJy~beam$^{-1}$.  In each case the beam size is illustrated in
the bottom left hand corner; and (b) 8.56~GHz with the VLA; contours
are $-$3, 3, 6, 12, 24, 48, 96 and 192 $\times$
80~$\mu$Jy~beam$^{-1}$, greyscale range is 0.08 to
18.96~mJy~beam$^{-1}$. (c) Extinction map (greyscale) derived from
Figs.~\ref{fig-HST1}(c) and \ref{fig-radio}(a), with VLA image at
23~GHz overlaid as contours.  Propagated uncertainty in E(B--V) is
$\sim$0.07. (d) Electron temperature versus density; solid lines are
the loci for values of $R_{OIII}$ as indicated. Dashed lines are for
$\nu$ = 8.56~GHz, $T_b$ = 140~K [lowest contour in
Fig.~\ref{fig-radio}(b)] and different values of $l$ (i) 0.00063~pc,
(ii) 0.005~pc, (iii) 0.025~pc.  Dotted lines are for $\nu$ = 8.56~GHz,
$T_b$ = 2260~K [fifth contour in Fig.~\ref{fig-radio}(b)] and
different values of $l$ (iv) 0.00063~pc, (v) 0.005~pc, (vi) 0.025~pc.
The dot--dashed line (vii) is for $\nu$ = 23~GHz, $T_b$ = 6760~K [the
peak of feature C in Fig.~\ref{fig-radio}(a)]. See text for
discussion.\label{fig-radio}\label{fig-compare}}
    \end{figure*}

\subsection{HST WFPC2 images}
\label{ssec-HST}

The {\em HST} WFPC2 images are shown in Figs.~\ref{fig-HST1} and
\ref{fig-HST2}. Some extended emission is apparent out to 2~arcsec,
but the brightest parts are within 0.5~arcsec of the central star,
which is readily identified from the diffraction spikes in
Fig.~\ref{fig-HST1}(c).  Feature~1 visible in Fig.~\ref{fig-HST1}(a)
is almost a ring of emission at the south--west edge of the inner
nebula. There are three compact knots to the east of the central
emission (features~2, 3 \& 4), and a fourth one to the south--west
(feature~5), which are most obvious in the F502N filter
[Fig.~\ref{fig-HST1}(a)].  Weak filaments are also present to the
north (feature~6 \& 7), and may be associated with northern
prominences seen at larger scales from the ground by
\citet{Corradi99}.  The inner parts of the nebula shows three
structures: a ridge to the north (N), a peak at the center (C) and a
more irregular feature to the south (S), which may be the brightest
part of a loop of emission.  In the case of the F656N image
Fig.~\ref{fig-HST2}(a), dominated by H$\alpha$, the central star
position has been fitted and the point spread function (PSF)
subtracted, using a PSF modeled using the TinyTIM software
\citep{Krist95}.  This image shows that features C, N and S are
distinct in hydrogen, and they are also present in the weaker H$\beta$
emission in the F487N image Fig.~\ref{fig-HST1}(c).  The features C, N
and S are present in the F218W image Fig.~\ref{fig-HST2}(d), but the
latter two features are extremely weak.

\subsection{VLA images}
\label{ssec-VLA}

The VLA images are presented in Figs.~\ref{fig-radio}(a) \& (b). At
23~GHz there is a northern ridge running E--W, a central, almost
circular peak, and a more irregular southern feature.  These features
coincide with features N, C and S in Fig.~\ref{fig-HST1}(a), and we
adopt the same labels for the radio structure.  From comparison of
Fig.~\ref{fig-radio}(a) and e.g. Fig.~\ref{fig-HST1}(a) we suggest
that at least one of the binary components is coincident with radio
feature~C.  We discuss this further in section~\ref{ssec-positions}.
The 8.56~GHz image [Fig.~\ref{fig-radio}(b)] has significantly lower
resolution, but is sensitive to more extended emission.  Two features
are apparent: a central N--S elongated structure, encompassing
features N, C and S, and an extension to the S--W coincident with
feature~1.

\subsection{Stellar positions}
\label{ssec-positions}

A simple estimate of the relative contributions of the HC and CC in
HM~Sge at a given wavelength can be made assuming blackbodies with
T$_{\rm eff}^{\rm HC}$ = 200~000~K \citep{Murset97}, T$_{\rm eff}^{\rm
CC}$ = 3~000~K, and taking approximate radii of 0.01 and
100~R$_{\odot}$ respectively.  This shows that the HC dominates the CC
by a factor $\sim~80$ at 220~nm [F218W, Fig~\ref{fig-HST2}(d)] and the
CC dominates the HC by a factor $\sim~18000$ at 550~nm [F547M,
Fig~\ref{fig-HST2}(b)].  Note that UV spectra show negligible line
contribution to the F218W filter \citep[see e.g.][]{Kenyon86}. A
number of lines contribute to the F547M filter \citep[see e.g.][their
Fig.~2]{Schmid00a}.  The CC is clearly badly modeled by this estimate,
as it does not account for the absorption bands characteristic of
Mira--types.  In addition, a simple estimate of the luminosity of the
HC \citet{Murset91} and the contribution at $\lambda\sim$2200\AA\
shows that the central pixel will be entirely dominated by stellar
emission, rather than nebular emission.  These estimates show that if
the two stars were displaced by more than a few pixels on the sky, the
peak of the emission should clearly move between the F218W image and
the F547M image.  That this is not the case means that both stars must
be within feature~C.

\begin{figure*}
%\epsscale{0.5}
%    \plotone{cartoon.eps} 
\begin{center}
\includegraphics[height=180mm]{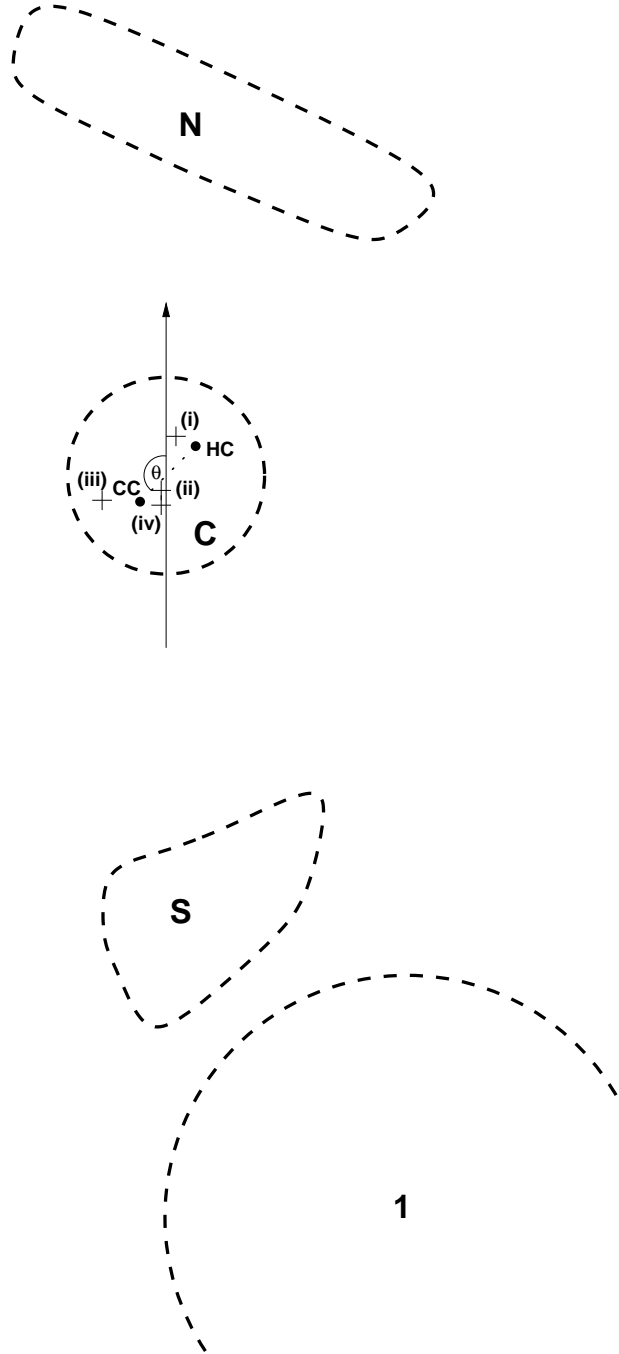}
\end{center}
\caption{Schematic illustrating the relationship
between the binary components and the features of the inner nebula
(not to scale.) North is up.  The position angle of the binary axis is
$\theta$=130$^\circ\pm$10$^\circ$.  The inner nebula features (N, C, S
and 1) are marked. The hot component (HC) and the cool component (CC)
are marked at the positions of the peak in the F218W and F547M image
respectively.  Other peak positions are (i) F437N, (ii) F469N (iii)
F487N \& F656N and (iv) F502N, marked as crosses.
\label{fig-cartoon}}
\end{figure*}

As the PSF of the star in each image extends to a large angular distance
from the peak, we have attempted to fit a Gaussian component to the
peak positions in the F218W and F547M images.  The best resolution was
available in our dithered images, where the pixel size is
0.02275~arcsec.  The position of the peak relative to 19~41~57
+16~44~39 in the F218W image [Fig.~\ref{fig-HST2}(d)] is
$\alpha$ = 0.08067$\pm$0.00004, $\delta$ = 0.6195$\pm$0.0005 while in the
F547M image [Fig.~\ref{fig-HST2}(b)] it is
$\alpha$ = 0.08258$\pm$0.000005 $\delta$ = 0.5920$\pm$0.00002.  The errors
quoted are the formal uncertainties of the fits.  When dithering, the
measured shifts (from the positions of bright stars in each frame) can
be up to 0.2 undithered pixels or $\sim$~9~milli--arcsec (mas)
different from the requested value of 5.5~pixels.  This is probably
the best estimate we have of the true uncertainty of the measured peak
shift from F218W to F547M.  Thus, the peaks in the two filters are
displaced from each other by 40$\pm$9~mas at a
position angle of 130$^{\circ}\pm$10$^{\circ}$ measured north through
east.  The position angle agrees well with that suggested by
\citet{Schmid00a}.  We suggest that the peak position in the F218W
image is that of the HC while the peak position in the F547M image is
that of the CC.  Thus, the line connecting the two stars on the sky,
the ``binary axis'', is at a position angle of
130$^{\circ}\pm$10$^{\circ}$.  The schematic in Fig.~\ref{fig-cartoon}
illustrates the relationship between the binary components and the
features apparent in the inner nebula.

To test further these measurements, we also looked at other
observations of symbiotic stars made under our GO programme.  In the
case of CH~Cyg (paper in preparation), the star was known to be in
eclipse during the observation, so that only one star could possibly
be seen.  The shift in that case was $\sim 10$~mas, comparable to the
estimate from the dithering uncertainties, and hence consistent with
the star being in eclipse.

The peak positions in the other filters are not necessarily coincident
with the stars as they are dominated by nebular lines.  These
positions are illustrated in Fig.~\ref{fig-cartoon}.  By definition
these filters are sensitive to more extended emission.  This means
that fitting a gaussian is invalid, as the nebulosity may be
asymmetrical but strongly emitting near the stars.  Thus a lack of
coincidence between the peaks of the narrow filters and the peaks of
the F218W and F547M filters is to be expected. The H$\alpha$ and
H$\beta$ images have peaks very close to that of the F547M image.  The
He~II image, which traces the highest ionisation regions, has a peak
between the two stars and slightly south of the binary axis,
consistent with the model of \citet{Nussbaumer90}.  Finally,
the F437N and F502N images, dominated by [O~III] lines which are
sensitive to temperatures and densities, have peaks which differ from
those of all the other images and from one another.  The results for
these latter three filters are most readily explained in the context
of the suggested wind interaction which might be expected to have the
strongest effects between the two stars.

\subsection{Nebular diagnostics}
\label{ssec-diagnostics}

A number of diagnostics of the physical conditions in the nebula can
be derived from the images presented here.  The ratio $R_{OIII}$ of
the F502N image (including both [O~III] $\lambda\lambda$4959 \& 5007)
to the F437N image (including [O~III] $\lambda$4363) depends on both
electron temperature $T_e$ and density $n_e$ according to the equation
\begin{equation}
\begin{array}{ll}
R_{OIII}  & \simeq \frac{j_{\lambda4959} +
j_{\lambda5007}}{j_{\lambda4363}}\\
& =
\frac{7.73exp[3.29 \times 10^4 T_e^{-1}]}{1 + 4.5 \times 10^{-4}(n_e
T_e^{0.5})}\end{array}
\label{eq-ROIII}
\end{equation}
\citep[see][]{Osterbrock89}.  Examination of ground--based optical
spectra indicate that these two WFPC2 filters are dominated by the
three lines which contribute to the ratio, and the nebular continuum
is very weak \citep[see e.g.][their Fig.~2]{Schmid00a}.  Thus,
Figs.~\ref{fig-HST1}(a) and (b) can be used to derive the ratio map
Fig.~\ref{fig-HST1}(d) which traces $T_e$ and $n_e$, and in principle
depends on both of these quantities.  This diagnostic has been applied
effectively to HM~Sge by \citet{Schmid90}, using spatially--unresolved
spectra for the entire nebula.

The WFPC2 Charge Transfer Efficiency (CTE) effect on the WFPC2 chips
\citep{Whitmore98a,Whitmore98b} has a bearing on this analysis.  For
aperture photometry, this causes stars at the top of the chip (row
number 800) to appear systematically fainter than those at the bottom
(row number 0).  The effect is also stronger for brighter stars.  The
consequences of this effect for extended emission is not well
understood. In our case the observed emission falls on the same region
of the chip for all observations, and brightness variations are not
severe across the nebula. This means that we can draw firm qualitative
conclusions from Fig.~\ref{fig-HST1}(d).

Radio brightness, expressed as a brightness temperature $T_b$, is
given by
\begin{equation}
T_b = T_e(1 - e^{-\tau_\nu}) ~~{\rm K},
\label{eq-Tb}
\end{equation}
where the optical depth at frequency $\nu$ is
\begin{equation}
\tau_\nu \simeq 8.24 \times 10^{-2} T_e^{-1.35} \nu^{-2.1} n_e^2 l
\label{eq-tau}
\end{equation}
assuming $n_e$ is constant, where $l$ is the path length through the
nebula \citep[again see][]{Osterbrock89}.  Thus, the radio brightness
temperature also depends on both $T_e$ and $n_e$.  Relationships
(\ref{eq-ROIII}), (\ref{eq-Tb}) and (\ref{eq-tau}) provide loci in a
$T_e$--$n_e$ plot, which intersect at the conditions appropriate to
the various features.  Comparing Figs.~\ref{fig-HST1}(d) and
\ref{fig-radio}(b), we can determine the conditions as a function of
position.  

Loci for typical values of $R_{OIII}$, $T_b$ and $l$ are illustrated
in Fig.~\ref{fig-compare}(d).  Also shown here is the locus for the
\citet{Schmid90} value of $R_{OIII} = 5.9$.  It is clear that the main
source of uncertainty in this analysis is the value adopted for $l$. A
reasonable estimate would be to place it at somewhere between the
largest and the smallest angular size seen on the sky.  When estimated
in this fashion, it also depends linearly on the distance, which is
not well known.  However, we should still be able to draw qualitative
conclusions about the relative temperatures and densities of the
different regions despite this distance uncertainty.  Most
importantly, it is apparent from Fig.~\ref{fig-compare}(d) that along
contours of constant $T_b$, Fig.~\ref{fig-HST1}(d) primarily traces
variations in $T_e$.  At the same time, variations in $T_b$ trace
variations in $n_e$. We must be careful not to use this analysis to
draw conclusions about those regions of the nebula where no radio
emission is detected.  An interferometer such as the VLA will {\em
resolve--out} smooth extended structure, meaning that the true
extended radio brightness will generally be greater than that present
in the radio map.  We note that the value of $R_{OIII}$ from
\citet{Schmid90} is effectively an average, dominated by the brightest
nebular feature.  We have plotted a locus in Fig~\ref{fig-compare}(d)
for the peak emission in the 23~GHz image (the peak of feature C) as
the most relevant one for comparison with the \citet{Schmid90} locus.
Given the variable nature of the inner nebula \citep{Richards99}, the
values which might be derived for that central region are similar to
those derived by \citet{Schmid90}, which were effectively a weighted
average for the entire nebula.

The F469N image [Fig.~\ref{fig-HST2}(c)] is dominated by He~II
$\lambda$4686 emission.  This is a discriminator of strongly ionized
regions \citep[see][]{Osterbrock89}.  It can be seen from the image
that features C, N and S are visible in He~II.  This suggests that
these features are strongly ionized, and is consistent with these
features being well defined in the F437N and F502N images
[Figs.~\ref{fig-HST1}(a) \& (b)], as [O~III] shows roughly the same
ionization structure.  In addition, the He~II emission does not extend
far beyond a radius of $\sim$1~arcsec, suggesting that the
intermediate scale nebulosity is more weakly ionized than the smallest
scale features.  This is more difficult to reconcile with the extended
structure seen particularly in Fig.~\ref{fig-HST1}(a).  However, as
noted above the [O~III] lines seen in these filters are sensitive to
temperature and density, so we may well be seeing the effects of
density inhomogeneities in the nebula.  Such inhomogeneities are in turn
consistent with the very clumpy nature of much of the emission at
sub--arcsec scales.

\subsection{Extinction mapping}
\label{ssec-extinction}

A third diagnostic comes from the fact that the {\em dereddenned}
H$\beta$ flux, $F({\rm H}\beta)$, can be derived from the radio flux
$S_\nu$ via the equation
\begin{equation}
S_\nu = 2.51 \times 10^{8} T_e^{0.53} \nu^{-0.1} F({\rm H}\beta) ~{\rm
Jy},
\label{eq-EB-V}
\end{equation}
where $F({\rm H}\beta)$ is in erg~cm$^{-2}$s$^{-1}$.  This assumes the
radio flux is due to optically thin thermal emission.  Thus, with a
measured H$\beta$ flux we can derive the reddening E(B--V).  In the
past this has been used to derive the interstellar reddening using the
total radio and H$\beta$ fluxes.  However, in dusty environments such
as that seen in HM~Sge, there may be a contribution to the reddening
from circumstellar matter. Such variations in circumstellar reddening
have been suggested to explain the different estimates of E(B--V) and
A$_V$ towards HM~Sge in the past \citep[e.g.][]{Munari89a}. In our
case, the images of both H$\beta$ emission [Fig.~\ref{fig-HST1}(c)]
and radio emission [e.g. Fig.~\ref{fig-radio}(a)] allow us to derive
an extinction map, as is shown in Fig.~\ref{fig-compare}(c) for $T_e$
= 10~000~K.  We note that \citet{Ivison92} present a
spatially--unresolved radio--infrared spectrum for HM~Sge, indicating
that the turn--over from optically thick to thin radio emission occurs
at 8.5~GHz.  However, it is clear from their Fig.~1 that the
turn--over is gradual, consistent with partially optically thin
emission up to $\sim$30~GHz.  In addition, the inner nebula of HM~Sge
is variable in both brightness and structure
\citep[e.g.][]{Richards99}, making the spectrum of \citet{Ivison92}
out of date.  This means that it remains difficult to assess the
optical depth conditions within the inner nebula as a function of
position.

The consequences of the CTE effect for Fig.~\ref{fig-compare}(c) are
to provide an overestimate of the extinction, with the largest over
estimates being at the brightest parts of Fig.~\ref{fig-HST1}(c).  This
extinction map is also affected by our assumptions about the radio
emission.  Equation~\ref{eq-EB-V} relies on the assumption that the
radio emission is optically thin thermal. While the brightness
temperatures are consistent with thermal emission, the optical depth
is not well constrained.  However, where emission is optically thick,
we would underestimate the extinction.  As the brightest radio
emission most likely comes from the regions of greatest optical depth, this
effect works in the opposite sense to the CTE effect.  Finally,
E(B--V) depends on electron temperature.  If $T_e$ = 40~000~K, instead
of the 10~000~K assumed above, E(B--V) would decrease by $\sim$~0.2.
Fig.~\ref{fig-compare}(d) demonstrates that $T_e$ can vary
considerably over the nebula.  However, for the larger values of $T_b$
in the inner regions relationships (\ref{eq-Tb}) and (\ref{eq-tau})
lead to loci at higher electron densities, while the curves derived
from equation (\ref{eq-ROIII}) converge towards lower $T_e$, and
extend over a narrower range of temperatures.  Thus in the inner
regions shown in Fig.~\ref{fig-compare}(c) the variations in $T_e$ are
relatively small.  As the hotter regions are closest to the white
dwarf, it seems likely that the intermediate values of E(B--V) = 0.7
may be somewhat reduced by the expected higher $T_e$ in that region.
Bearing in mind these considerations, we are confident that we can
draw qualitative conclusions from Fig.~\ref{fig-compare}(c).

\section{Discussion}
\label{sec-disc}

\subsection{Dust distribution and interstellar extinction}
\label{ssec-dust}

The extinction map in Fig.~\ref{fig-compare}(c) can be taken to trace
the circumstellar variation in the dust distribution, assuming the
interstellar extinction is reasonably uniform over the relatively
small angular size of the inner nebula.  In this case, the best
estimate of the true interstellar extinction comes from the minimum
values in the extinction map.  Three distinct regions are apparent:
(i) to the south E(B--V) $\simeq$ 1, coincident with feature~S; (ii)
to the north around feature~N, E(B--V) $\simeq$ 0.6, consistent with
previous reddening estimates; and (iii) a low--ratio band running from
east to west across the position of feature~C. Note that the peak
itself does not give a reliable E(B--V) estimate, as the emission in
the optical is not entirely nebular.  This band is also weakly
contaminated by the diffraction spikes, which have a position angle of
$\sim 80^{\circ}$.  The eastern part of this low--ratio band has
E(B--V) = 0.35, the lowest value on the map, and presumably the best
estimate of the true interstellar extinction.  As high E(B--V)
correlates with a greater quantity of dust, we can see that the dust
is concentrated at feature~S.  In addition, there is a deficit of dust
closer to feature~C in Fig.~\ref{fig-HST1}(a), especially immediately
to the east.  This clumping of the dust at small scales was suggested
by \citet{Thronson81}, based on discrepancies in reddening estimates.
Our maximum value of E(B--V)~$\simeq$~1 gives A$_V~\simeq$~3.1 for
interstellar dust \citep{Evans94}, which is much lower than the value
of A$_V \simeq$ 12 of \citet{Thronson81} but similar to other
estimates both before and after that time. Much more recently,
spectro--polarimetry by \citet{Schmid00a} demonstrated that the Mira
was clearly visible, in agreement with the significant reduction in
A$_V$ since 1980 we see here.  Results from the Infrared Space
Observatory (Schild et al., submitted) are consistent with a
two--component dust model, with optical depths comparable to those
derivable from our E(B--V) values.  Such a model is also consistent
with the extinction map in Fig~\ref{fig-compare}(c), particularly as
the Schild et al. modeling suggests that at least one of the
components is clumpy.

We suggest that the highest concentration of dust, at feature~S, is
associated with the CC wind.  Note that Fig.~\ref{fig-compare}(c) is
blanked beyond the 1$\sigma$ level of the VLA image, as no useful
information can be derived beyond this limit.  Hence, we have no
knowledge about the dust distribution beyond that region.

\subsection{The distance and the binary parameters}
\label{ssec-distance}

The minimum in the extinction map Fig.~\ref{fig-compare}(c) places the
interstellar extinction at no more than E(B--V) = 0.35.  This is an
upper limit because there may be a circumstellar contribution to the
minimum.  This gives an interstellar total extinction $A_V \leq$1.1
magnitudes.  \citet{Neckel80} provide maps of the galactic extinction
versus distance.  HM~Sge lies at the edge of their field 265 (see
Figs.~5 and 6m of \citealt{Neckel80}).  For $A_V$ = 1.1, the minimum
distance to HM~Sge is approximately 700~pc, although measurements of
similar $A_V$ are shown in field~265 for distances as low as 400~pc
and as high as 2~kpc.  Comparison with the extinction maps of
\citet{Lucke78} suggest a distance of $\sim$800~pc for this value of
E(B--V).  However, such large--scale extinction maps are based on a
small number of lines of sight in each field, so that the
distance--dependent extinction values are not readily applicable to
individual objects for the purposes of distance estimates.

Our measured binary angular separation (section~\ref{ssec-positions})
suggests a projected binary separation of (40$\pm$9)D~au at a distance
of D~kpc.  This can be reconciled with the results of
\citet{Richards99} if the distance to HM~Sge is
$\sim$~1250$\pm$280~pc, which is consistent with our estimates above.

\subsection{Nebular temperatures}

The map of $R_{OIII}$ [Fig.~\ref{fig-HST1}(d)] shows two distinct
regions in the extended nebula visible with the {\em HST}.  Most of
the nebula has 4 $< R_{OIII} <$ 15, while a wedge region to the
south--west has $R_{OIII} >$ 15 [coincident with feature~1 in
Fig.~\ref{fig-HST1}(a)].  At the center, where the stars contribute
significantly, $R_{OIII}$ is as low as 2, but clearly this is not
purely nebular emission, and so cannot be relied upon for nebular
diagnostics.  From Fig.~\ref{fig-compare}(d) it is clear that higher
$R_{OIII}$ regions are consistent with lower electron temperatures
along a given $T_b$ contour.  This wedge extends from roughly the
location of the highest extinction in Fig.~\ref{fig-compare}(c) to the
edge of the nebula, and is suggestive of the ``bow wave'' structures
seen in interacting--wind simulations.  More importantly, there is a
cooler region in the nebula associated with the southern dust feature,
further supporting our suggestion that we are seeing the CC wind
directly. Alternatively, this region of the nebula could be shielded
from the HC radiation field by feature~S.

\section{Conclusion}
\label{sec-conc}

Using contemporaneous {\em HST} WFPC2 and VLA observations of the
symbiotic nova HM~Sge, we have investigated the nebular conditions and
the relationship of the star positions to the nebular structure.  We
identify a number of discreet features at spatial scales smaller than
$\sim$~0.1~arcsec embedded in the extended nebula, including at least
four discreet knots, and two linear features to the north--east.  A
ring--like feature (labeled~1) is evident on the south--west edge of
the inner nebula.

The radio and optical emission is well correlated in the inner
1~arcsec, with a northern ridge (N), a central peak (C), and a more
irregular southern feature (S) clearly evident at both 23~GHz and
$\sim$5000\AA.  Temperature diagnostics show a ``wedge'' to the
south--west, encompassing feature~1, consistent with lower
temperatures than the rest of the nebula. The knots 1--3 and the
western tip of ridge N also appear to have cooler, denser conditions
than the rest of the nebula, but these are less well defined in the
diagnostic map.

Both stars must be encompassed by feature~C.  The peak shifts by
40$\pm$9~mas between the F218W image ($\sim$~220~nm) and the F547M
image ($\sim$~550~nm), with a position angle of
130$^{\circ}\pm$10$^{\circ}$.  We place the HC at the continuum UV
peak and the CC at the red continuum peak.  The position angle of the
line between the two stars on the sky, the ``binary axis'', is
130$^{\circ}\pm$10$^{\circ}$. This is consistent with a suggested
binary axis from spectro--polarimetry \citep{Schmid00a}.  It is also
consistent with a suggested binary separation of 50~au
\citep{Richards99} if the distance is $\sim$~1250$\pm$280~pc.

The circumstellar extinction map suggests a patchy dust distribution,
with the greatest extinction to the south--west, coincident with
feature~S, and the ``wedge'' around feature~1.  The minimum in the
extinction map indicates the true interstellar extinction to be no
more than E(B--V) = 0.35.  This is consistent with a minimum distance
of $\sim$700~pc, but this would be reduced if there is a circumstellar
contribution to the minimum. We suggest feature~S and the south--west
wedge (feature~1) are associated with the CC wind.  Alternatively,
the south--west wedge is shielded from the HC radiation field by
feature~S.

\acknowledgments

\noindent Thanks are due to Dr S. J. Smartt of the UK {\em HST}
Support Unit for his essential assistance with the details of the
imaging.  SPSE is supported by a Research Assistantship Award from the
Particle Physics and Astronomy Research Council (PPARC). MMC is
supported by a PhD Studentship Award from PPARC.  The VLA is operated
by the National Radio Astronomy Observatory, a facility of the
National Science Foundation operated under cooperative agreement by
Associated Universities, Inc.

% \figcaption[HST1.eps]{}

%\figcaption[HST2.eps]{}

%\figcaption[radio.eps]{}

%\figcaption[cartoon.eps]{}

\end{document}